\documentclass{iau-JDSS}

\usepackage{graphicx}
\usepackage{natbib}

\bibpunct[, ]{(}{)}{;}{a}{}{,}

\title[Pushing the limit of instrument capabilities] 
{Pushing the limit of instrument capabilities}

\author[D. Shulyak, W. Weiss, G. Mathys, L. Eyer, A. Kholtygin, et al.]   
{Denis V. Shulyak$^{1}$, Werner W. Weiss$^2$, Gautier Mathys$^3$, Laurent Eyer$^4$, Alexander F. Kholtygin$^5$,
Oleg Kochukhov$^6$, Pierre North$^7$, Sergey N. Fabrika$^8$, Tatyana E. Burlakova$^8$}

\affiliation{$^1$Institute of Astrophysics, Georg-August-University,\\
Friedrich-Hund-Platz 1, D-37077 G\"ottingen, Germany\\
email: {\tt denis@astro.physik.uni-goettingen.de}\\[\affilskip]
$^2$Institut f\"ur Astronomie, Universit\"at Wien,\\
T\"urkenschanzstra{\ss}e 17, 1180 Wien, Austria\\
email: {\tt werner.weiss@univie.ac.at}\\[\affilskip]
$^3$European Southern Observatory,\\
Casilla 19001, Santiago 19, Chile\\
email: {\tt gmathys@eso.org}\\[\affilskip]
$^4$Observatoire de Gen\'ve,\\
51 ch. des Maillettes, CH-1290 Sauverny, Switzerland\\
email: {\tt Laurent.Eyer@unige.ch}\\[\affilskip]
$^5$Astronomical Institute, St. Petersburg State University,\\
Universitetskii pr. 28, St. Petersburg, 198504, Russia\\
email: {\tt afkholtygin@gmail.com}\\[\affilskip]
$^6$Department of Physics and Astronomy, Uppsala University,\\
Box 515, 751 20, Uppsala, Sweden\\
email: {\tt Oleg.Kochukhov@fysast.uu.se}\\[\affilskip]
$^7$Ecole Polytechnique F\'ed\'erale de Lausanne,\\
1015 Lausanne, Switzerland\\
email: {\tt pierre.north@epfl.ch}\\[\affilskip]
$^8$Special Astrophysical Observatory, Russian Academy of Sciences,\\
Nizhnii Arkhyz, Karachai Cherkess Republic, 369167, Russia
}

\pubyear{2009}
\volume{15}  
\pagerange{1--10}
\jname{Highlights of Astronomy}
\editors{Ian F Corbett, ed.}

\begin{document}

\maketitle

\begin{abstract}
Chemically Peculiar (CP) stars have been subject of systematic research since
more than $50$ years. With the discovery of pulsation of some of the cool
CP stars, the availability of advanced spectropolarimetric instrumentation and
high signal-to-noise, high resolution spectroscopy, a new era
of CP star research emerged about $20$ years ago. 
Together with the success in ground-based observations, 
new space projects are developed that will greatly benefit
for future investigations of these unique objects. In this contribution we will give an overview of
some interesting results obtained recently from ground-based observations
and discuss on future outstanding Gaia space mission and its impact on CP star research.
\end{abstract}

\firstsection 
\section{Brief overview on CP stars}

Back in $1897$, more than one century ago, the first peculiar stars were found 
in the course of the Henry Draper Memorial classification work at
Harvard by Antonia Maury and Annie Cannon. Maury used the designation ``peculiar''
for the first time to describe spectral features in the remarks to the spectrum of
$\alpha^2$~CVn \citep{pickering1897}, making a first attempt for two-dimensional
classification system considering the strength and the width of the
spectral lines.

In $1974$ Preston proposed the division of main-sequence CP stars into four groups according to their spectroscopic
characteristics \citep{preston}: CP1 (Am/Fm stars), CP2 (Si, SrCrEu stars),
CP3 (HgMn stars), CP4 (He-weak stars). More detailed spectroscopic consideretion of CP stars
required to introduce new subtypes of CP stars, such as He-rich and $\lambda$~Boo stars. 
CP2 stars, including Bp/Ap, host strong
surface magnetic fields \citep{lantz} that are likely stable on large time-intervals \citep{north}.


Abundance peculiarities were measured using the curve-of-growth method
based on simple assumptions about formation of absorption lines (models of Schuster-Schwarzschild, Uns\"old, Milne-Eddington).
Since that time and with development of new high-resolution, high signal-to-noise CCD based spectrometers, big
progress has been made in abundance analysis of CP stars, revealing the presence of vertical \citep{str1,str2,str3,str4} 
and horizontal \citep{di0,di1,di2} elements separation in their atmospheres caused by the processes of
microscopic particle diffusion \citep{michaud}.

The discovery of strong stellar surface magnetic fields \citep{babcock} opened a new research 
field in astrophysics~--~stellar magnetism. A $200$~Gauss accuracy of the magnetic field detection usually
obtained with photographic plates has increased to $\approx1$~Gauss with modern spectropolarimetry
and new techniques (such as Least Square Deconvolution, or LSD, for example) \citep{lsd,wade}.

With the discovery by D.~Kurtz \citep{kurtz1978} of a $12$~min pulsation period in HD\,101065 a subgroup of the cool 
CP stars, the so-called rapidly oscillating Ap (roAp) stars, became extremely promising targets 
for asteroseismology, a most powerful tool for testing theories of stellar structure. 
Driving of the oscillations results from a subtle energy balance depending directly on the 
interaction between the magnetic field, convection, pulsations, and atomic diffusion. 
Amazing insights in the 3-D structure of stellar atmospheres became available 
\citep[see, for example,][]{puls1,puls2}.

\section{Selected results from recent ground-based observations of CP stars}

Current ground-based observations of CP stars reveal many interesting findings that require new modelling
approaches and interpretations.

Accurate high-resolution observations are needed to understand the properties and origin of so-called hyper-velocity stars
(HVS). These are B-type stars with peculiar galactic rest frame velocity 
and enhanced $\alpha$~elements and normal (solar) Fe abundances in their atmospheres.
The origin of these stars is not well understood: one of the hypothesis tells that they originate from the dynamical 
interaction of binary stars with the supermassive black hole in the Galactic Centre (GC), which accelerates one 
component of the binary to beyond the Galactic escape velocity. So far, however, no HVS has been unambiguously related to a
GC origin. Determination of the place of ejection of a HVS requires the determination of space motion (accurate proper motions)
and chemical composition, however, the later one is hard to use to constrain their origin \citep[see][]{przybilla}. 
On the other hand, if GC origin can be proved by such future astrometric missions like Gaia (see below in this contribution) 
then it can bring important constraints on magnetic fields in GC region and on the formation and evolution 
of CP stars in general.

Important issue concerns a rotational braking observed recently for magnetic Bp star HD~37776. It was shown that this
star increased its rotational period by $17.7$s over the past $31$ years \citep{hd37776}. This can not be explained
by light-time effect caused, for example, by the presence of a secondary companion which is not observed in radial velocity
measurements. Also, the hypothesis of free-body precession due to magnetic distortion is incompatible with light curve 
shapes unchanged in $31$ years. The plausible scenario left is a continuous momentum loss due to magnetic braking that
requires magnetically confined stellar wind, which naturally can be present in HD~37776.

Interferometry is becoming a very powerful tool in modern astrophysics since it allows for direct and model independent
measurement of sizes of stellar objects. Recently, there was a report of the first detailed interferometric study of a
roAp star $\alpha$~Cir \citep{acirb}. Authors used observations of Sydney University Stellar Interferometer (SUSI) and 
additional data from visual and UV observations calibrated to absolute units to accurately derive $T_{\rm eff}$ and $\log(g)$
of the star. Even thought the interferometry can provide us with accurate radii of stars, 
the accuracy of $T_{\rm eff}$ determination is still limited by the incompleteness of observations in all spectral
range (that would give a value of the bolometric flux) and model atmospheres used \citep[see][]{acir}.
Implementation of interferometric techniques to another Ap star $\beta$~CrB is reported in this conference (JD04-p:29).

Interesting results were obtained for well known Ap star $\epsilon$~UMa which appeared to host a brown dwarf companion.
This star shows substantial variations of radial velocities measured from different spectral lines \citep{Woszczyk1980}
which allowed to derive the parameters of the system and to infer the mass of the secondary component of
$M_{\rm 2}=14.7M_{\rm Jupiter}$ \citep{sokolov2008}.

There have been an extensive discussions over past years about the LiI~$6708$\AA\ 
line which is usually used for the determination
of Li abundance in magnetic Ap stars. However, due to unknown and yet unconfirmed blending with possible lines of rare-earth
elements (REE) it was not definitely clear if the abundance determination suffer from systematic uncertainties or not.
Recently \citet{li} confirmed Li identification in magnetic Ap stars using detailed calculations of Zeeman pattern
in Baschen-Back regime.
No correlation between Li and REE line strengths was found thus ruling out the suspicion that the observed feature 
is due to an unidentified REE line. However the origin of Li still has to be explained theoretically by
diffusion calculations and/or abundance spots on stellar surface.

High-resolution, high signal-to-noise observations allow to study not only elements stratification in atmospheres 
of CP stars with very detail, but also the separation of different isotopes of a given element as demonstrated
by \citet{cowley} who studied $^{40}$Ca/$^{48}$Ca isotopic anomaly in atmosphere of selected Ap stars. This was
also investigated in \citet{ryabchikova2008} who analysed calcium isotopes stratification profiles
in three stars $10$~Aql, HR~1217 and HD~122970 concluding that the heavy isotope concentrated towards the higher layers.
Interestingly, they found no correlation in $^{48}$Ca excess in atmospheres of roAp and noAp stars with the magnetic field strength.

\begin{figure}[ht!]
\centering 
 \includegraphics[height=1.35in,width=4.95in,angle=0]{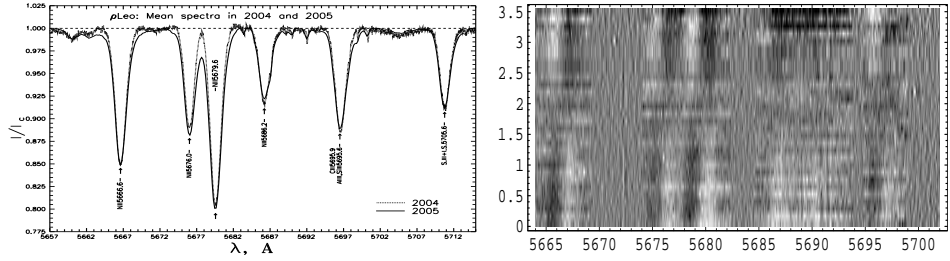}
 \caption{\small {\bf Left Panel}: Mean spectra of $\rho\,\,$Leo in the spectral region 
               $5667-5715\,$\AA\ in 2004 and 2005 years. {\bf Right Panel}: {\it dynamical} spectra of lpv 
               in the same region. 
          }
\label{Fig.CompareDist}
\end{figure}

New results of searching for the line profile variability (LPV) in the spectra of 
OB stars have been recently reported based on observations made with the $1.8$-m telescope of Korean Bohyunsan
Optical Astronomical Observatory and $6$-m telescopes of Special Astrophysical Observatory, Russia. 
For all program stars the regular and often coherent for all spectral lines LPV was reported,
as shown in Fig.~\ref{Fig.CompareDist} (right panel) for the B1 supergiant $\rho\,$Leo. 
This coherence is connected with the presence of the stellar magnetic field. 
The moderate dipole magnetic field of $\rho\,\,$Leo with the polar field strength $B_p\approx250$~G 
was detected by \citet{Kholtygin-2007}.

Together with the regular LPV the numerous local details
of line profiles in spectra of a few program stars are detected and are connected
with the formation and destruction of the small-scale 
structures (clumps or clouds) in the stellar wind. The evidence that numerous clumps exist in 
the winds of the O6 star $\lambda$~Ori~A and O9.5 star $\delta$~Ori~A was 
found. The {\it dynamical wavelet spectra} of LPV technique is used \citep{Kholtygin2008} 
to determine the distribution of the line fluxes for the clump ensemble in the winds of these stars. 

Many new interesting results have been obtained in the past couple of years from ground-based observations of CP stars,
that lead to refinements of ``traditional'' analyses based on observations of ever increasing quality and on more
refined analyses evidence effects that had not been previously detected. Definitely, ``new'' observational techniques 
(such as interferometry) have a large potential. Additional observational constraints will
force us to re-think some of our ``well-established'' views on the nature, origin and evolution of CP stars
bringing bridges between them and some topics of present-day astrophysics (galactic center, lowest-mass stars, etc.).

\section{Gaia space mission and CP stars}

In parallel to the progress in ground-based observations, modern space missions open a number of possibilities for
CP stars research not only in solar neighbourhood, but also much far beyond. Scheduled to be launched in 2012,
Gaia is one of the cornerstone astrometric mission of European Space Agency (ESA)
which will allow to study general properties and characteristics of a huge sample of stars of all spectral types.
Gaia will measure position, distances, space motions, radial velocities and fundamental parameters 
of about $1$~billion(!) of sky-objects. 
This scanning satellite will observe
every single object (galaxies, quasars, solar system objects, etc.) up to $20$th magnitude, 
thus providing important scientific constrains almost for all fields of modern astrophysics. With the expected
astrometric accuracy of about $7\mu$as at $10$th magnitude, Gaia's precision is more than $100$ times higher than those
of previously developed missions like Tycho and Hipparcos. Gaia has two telescopes with two viewing directions
separated by $106.5^\circ$: each telescope has SiC primary mirror $1.45\times0.5$~m$^2$ and $35$~m focal length.
The key idea of Gaia design is that the images from both telescopes are combined in one focal plane which hosts a number of
CCD detectors with the total size of $4500\times1966$ pixels. Many interesting and much more detailed information
can be retrieved from official Gaia website (http://www.rssd.esa.int/index.php?project=GAIA\&page=index).

Although Gaia is at first an astrometric mission, it will also provide us with broad-band spectrophotometric data in
wide spectral range. This is done by two photometers called Blue Photometer (BP) and Red Photometer (RP) with the
working wavelength's ranges of $330-680$~nm and $640-1000$~nm respectively. In addition, Gaia will be armed with spectrograph
for the radial velocity measurements (RVS) which operates in a narrow spectral range of $847-874$~nm 
(around CaII infrared triplet) with the resolution of $R=11500$. Every detected object will automatically 
pass through BP/RP and then RVS CCD's, and this opens a wide range of possibilities for stellar studies. 
Indeed, RP/BP photometry and RVS spectroscopy will ideally allow the determination of such important parameters
as $T_{\rm eff}, \log(g)$, and metallicity based on specific calibrations and sophisticated algorithms that are
presently developed: the final catalog of Gaia observations is scheduled to 2020, after 5 years of operation and 
three years of data reduction phase.

\begin{figure}
\begin{center}
 \includegraphics[height=4cm]{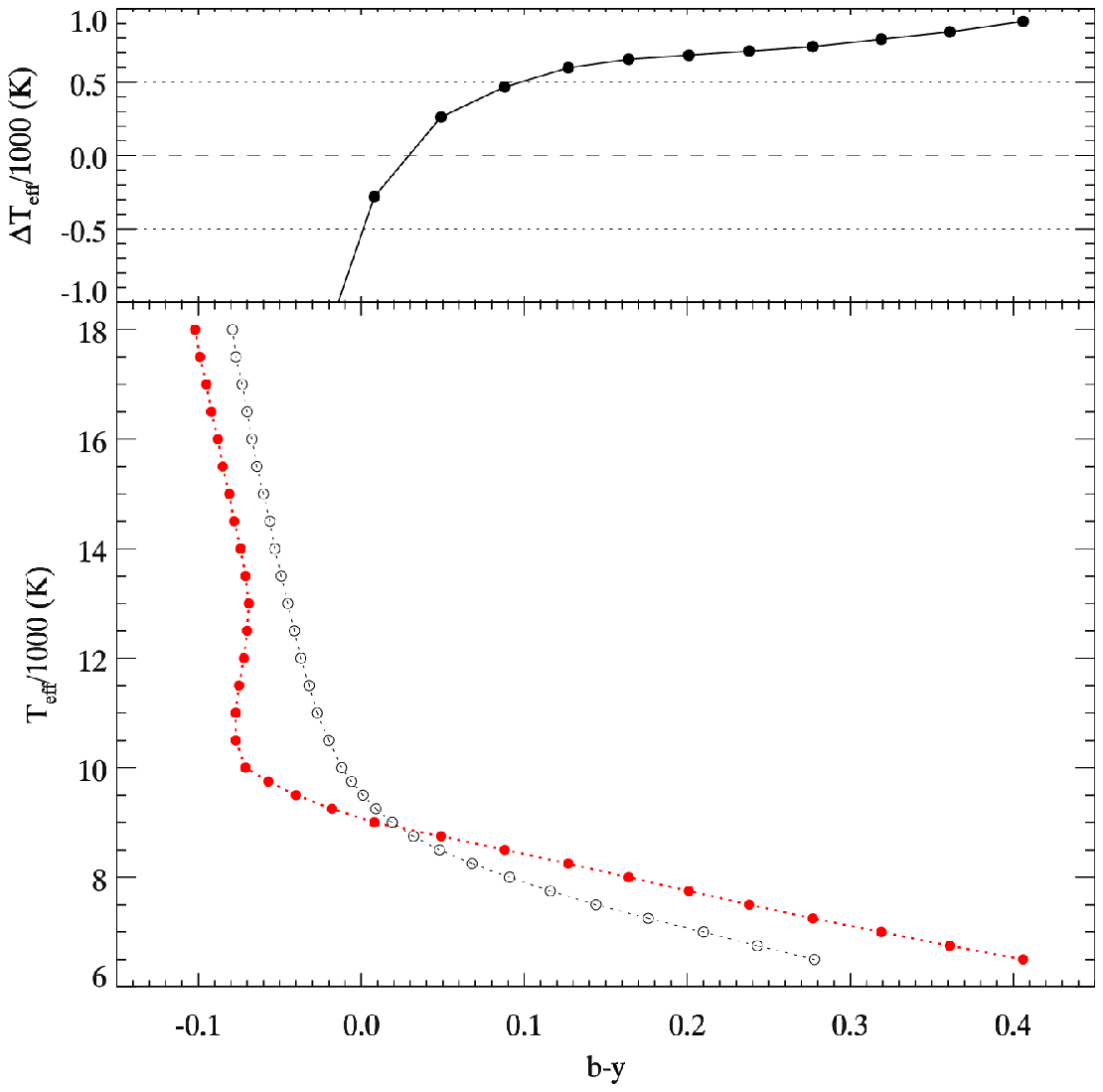}\hspace{0.5cm}
 \includegraphics[height=4cm]{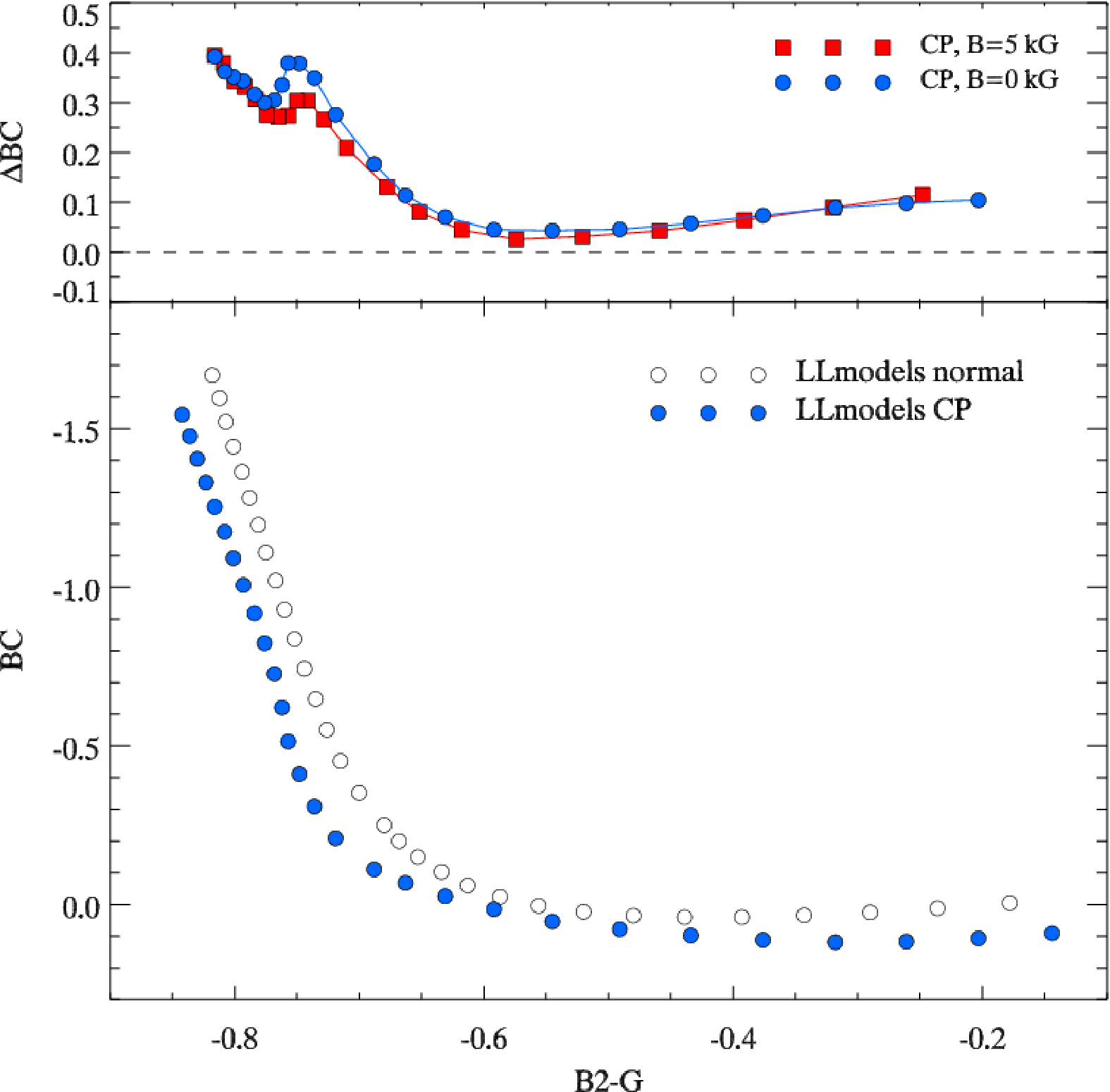}
 \caption{\textit{Left panel}: theoretical $T_{\rm eff}$ as a function of Str\"omgren $b-y$ index for normal (open cycles)
and CP (filled cycles) stars. \textit{Right panel}: same but for bolometric correction BC as a function of $B_{\rm 2}-G$
Gen\'eva index. Upper subplot shows difference in BC of non-magnetic CP and $5$~kG magnetic CP models with respect to models
computed with solar abundances.}
 \label{fig:peculiar}
\end{center}
\end{figure}

Among $1$~billion of objects, there will be a certain fraction of CP stars observed. Indeed, depending on spectral
classes, $10-30$\% of main-sequence B-F stars are CP, and at A0 all slow rotating stars are CP. The determination of
the position of these stars, their parameters, binarity, class of peculiarity etc., would greatly benefit to our understanding
of CP phenomenon in general and increase the number of potentially interesting objects for detailed research in particular.
However, we are faced a difficulty that for stars with moderate or strong peculiarities the 
standard temperature indicators are inadequate and blind application of usual photometric calibrations may lead
up to $500-1000$~K errors in $T_{\rm eff}$ determination \citep[see, for example,][]{hd137509}. 
Generally, there are two reasons why the energy distributions
of CP stars differ from those of normal: inhomogeneous horizontal and vertical elements distribution as well as 
the presence of strong surface magnetic fields. All these modifies atmospheric structure and thus lead to abnormal 
photometric parameters observed for CP stars. As an example, the impact of peculiar opacity on some photometric 
parameters are shown in Fig.~\ref{fig:peculiar}, which illustrates theoretically predicted behaviour of some
parameters. Theoretical models were computed taking into account characteristic 
temperature behaviour of abundances of CP stars as derived, for example, in \citet{ryabchikova2004}.

Thus, in order to distinguish between normal and CP stars as observed by Gaia, as well as to derive the type of peculiarity
for individual objects, a parametrization of CP stars is needed. This, in turn, requires dedicated model atmospheres to predict
their observed parameters. In out investigations we employ the 
\textsc{LLmodels} stellar model atmosphere code to compute extensive libraries of high resolution stellar fluxes that are now 
used for preliminary analysis of Gaia simulated data. \textsc{LLmodels} is 1-D, LTE model atmosphere code \citep{llm} that
treats the bound-bound opacity by direct, line-by-line spectrum synthesis with the fine frequency spacing 
($10^5-10^6$ points, $10^7$ in parallel mode) thus resolving individual spectral lines. The code does not use any precalculated
opacity tables and no assumptions are made about the depth-dependence of the line absorption coefficient thus providing a high dynamical
range in opacity calculation. This makes it possible to account for the effects of individual 
non-solar abundance and inhomogeneous vertical distribution of elements. It can also compute models with detailed treatment
of anomalous Zeeman splitting \citep{zeeman_paper1}, polarized radiative transfer in all four Stokes parameters \citep{zeeman_paper2,zeeman_paper3}, 
and magnetohydrostatic equilibrium taken into account \citep{lorentz,lorentz2}.
Chemical abundances and stratification are provided as input parameters for the
\textsc{LLmodels} code and kept constant in the model atmosphere calculation process. 
This allow us to explore the changes in model structure
due to stratification that were extracted directly from observations without modelling the processes that could be responsible
for the observed inhomogeneities. Such an empirical modelling can be applied to any CP star for which accurate spectroscopic
observations exist \citep[see][and talk JD04-i:5 of this meeting]{acir,hd24712}.

\begin{figure}
\begin{center}
 \includegraphics[height=4cm]{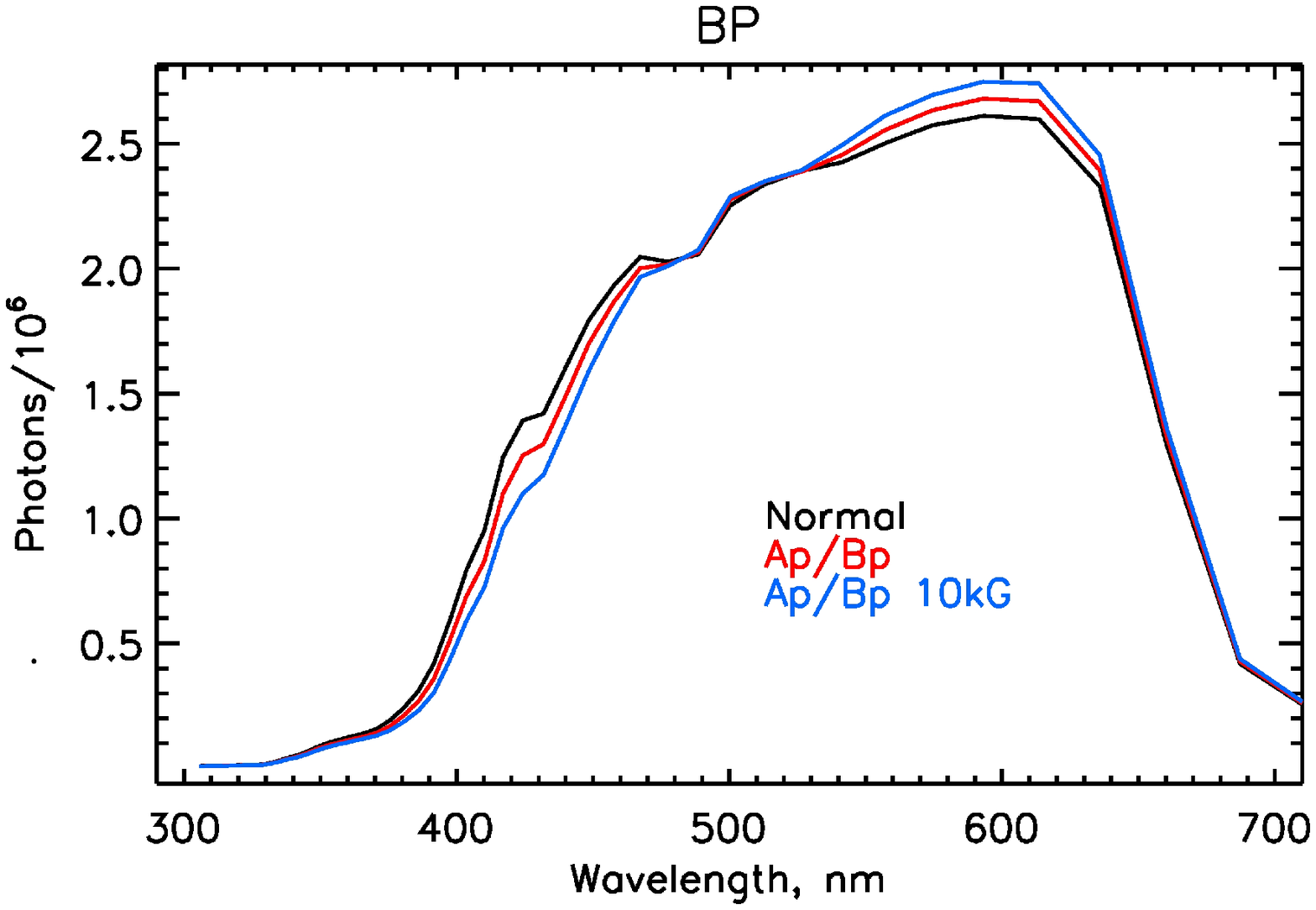}
 \includegraphics[height=4cm]{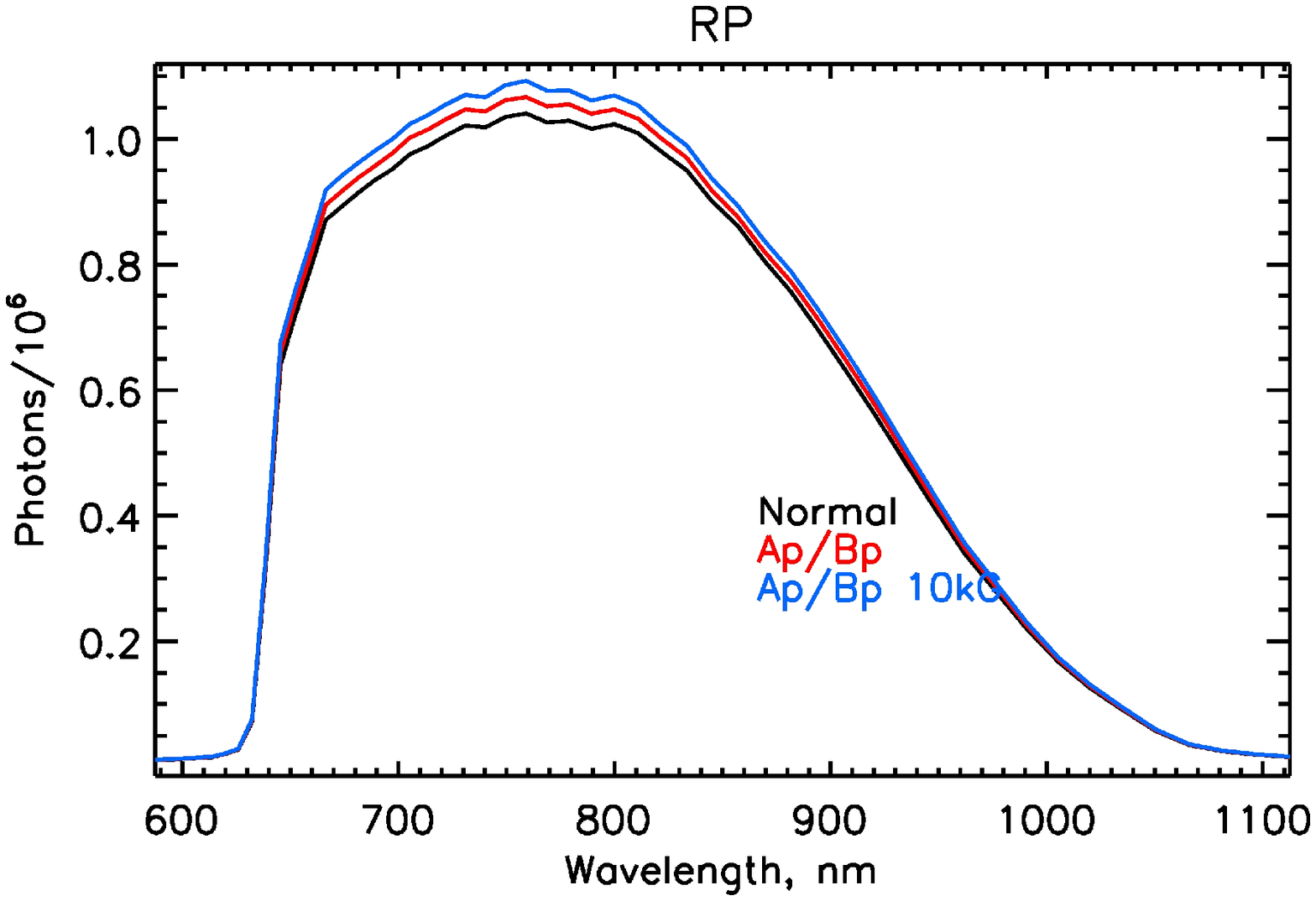}
 \includegraphics[height=3.5cm]{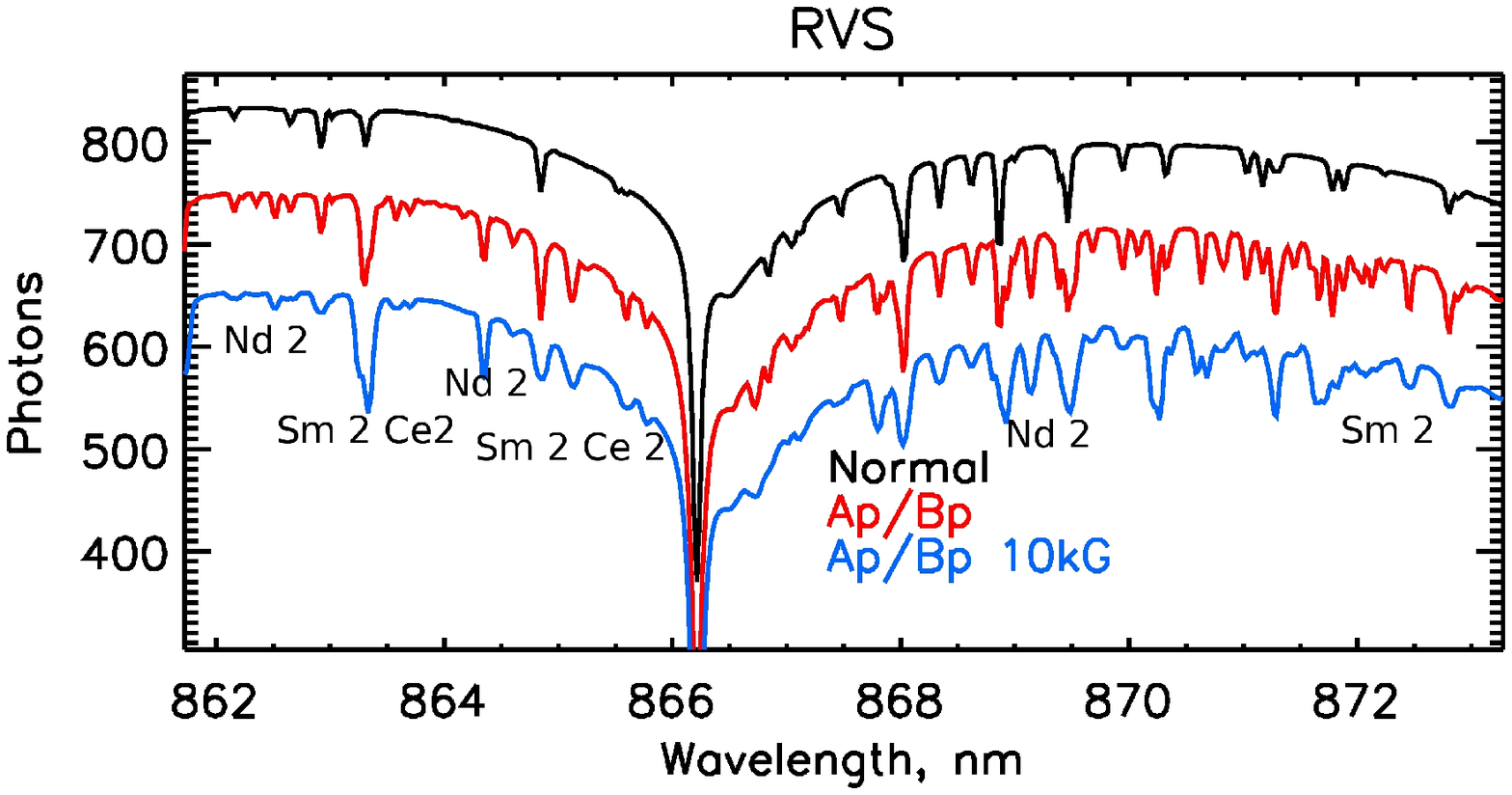}
 \caption{Gaia simulated BP/RP and RVS data for $T_{\rm eff}=8000$~K, $\log(g)=4.0$ normal and magnetic and non-magnetic Ap stars.}
 \label{fig:ap}
\end{center}
\end{figure}

\begin{figure}
\begin{center}
 \includegraphics[height=4cm]{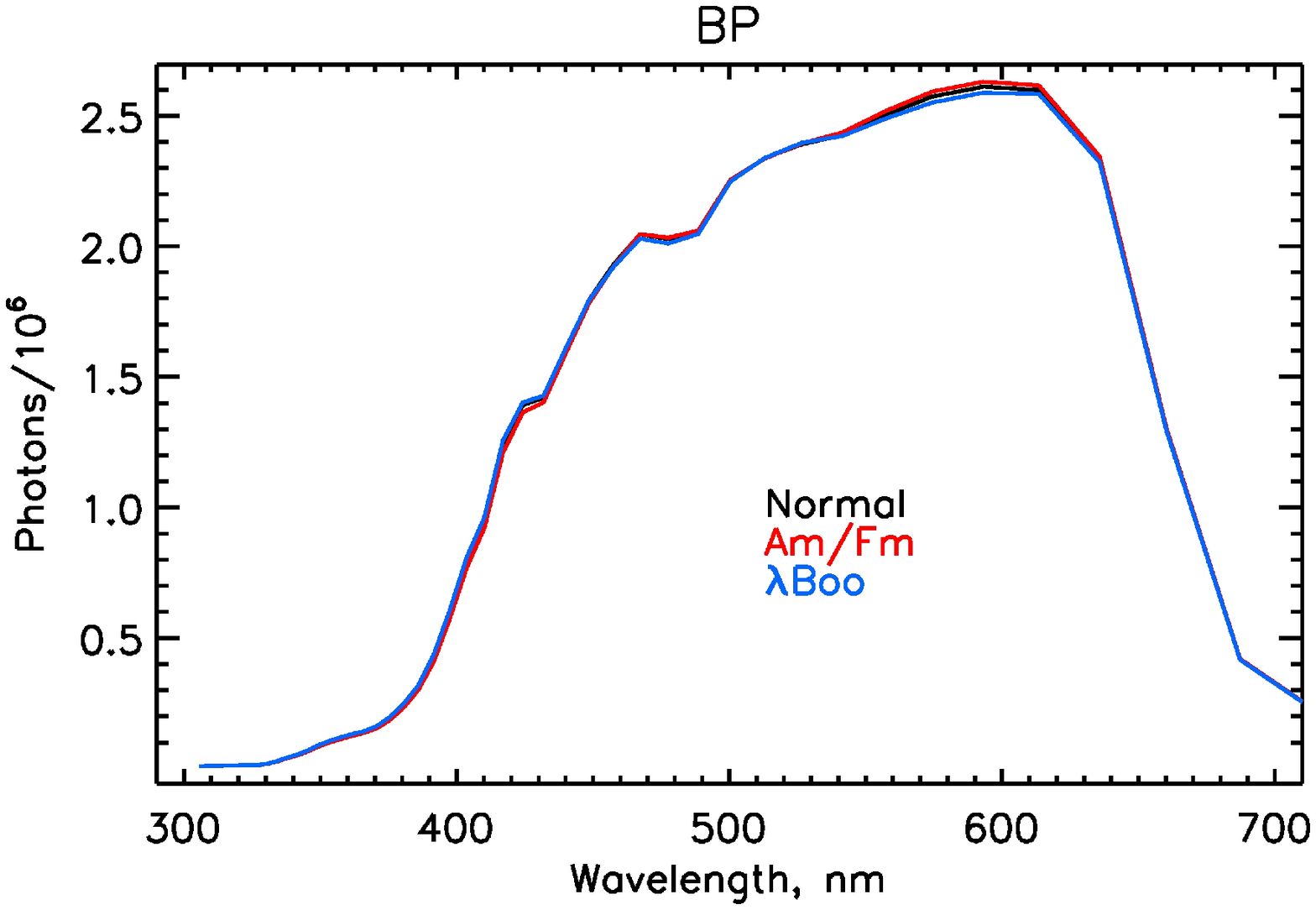}
 \includegraphics[height=4cm]{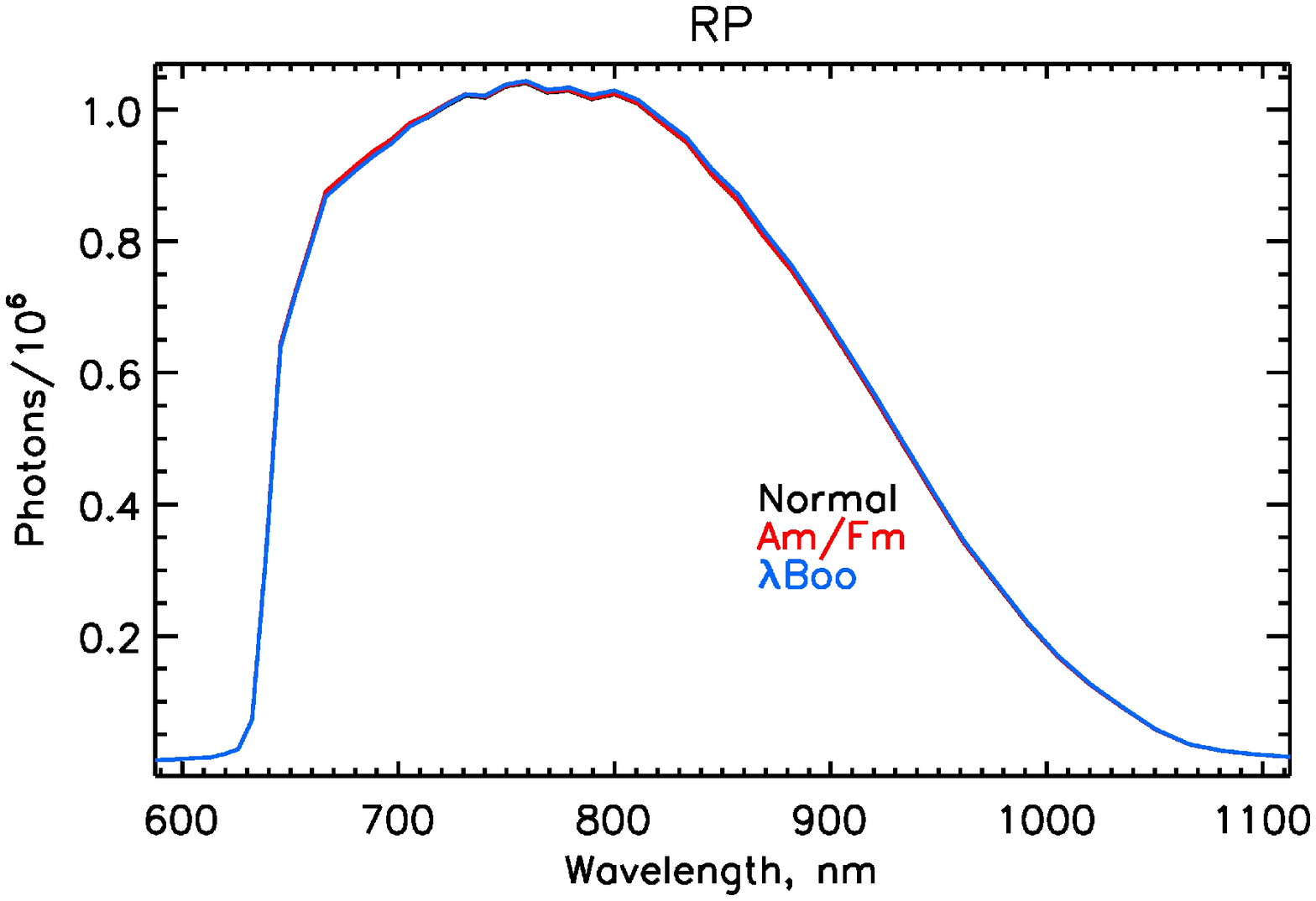}
 \includegraphics[height=3.5cm]{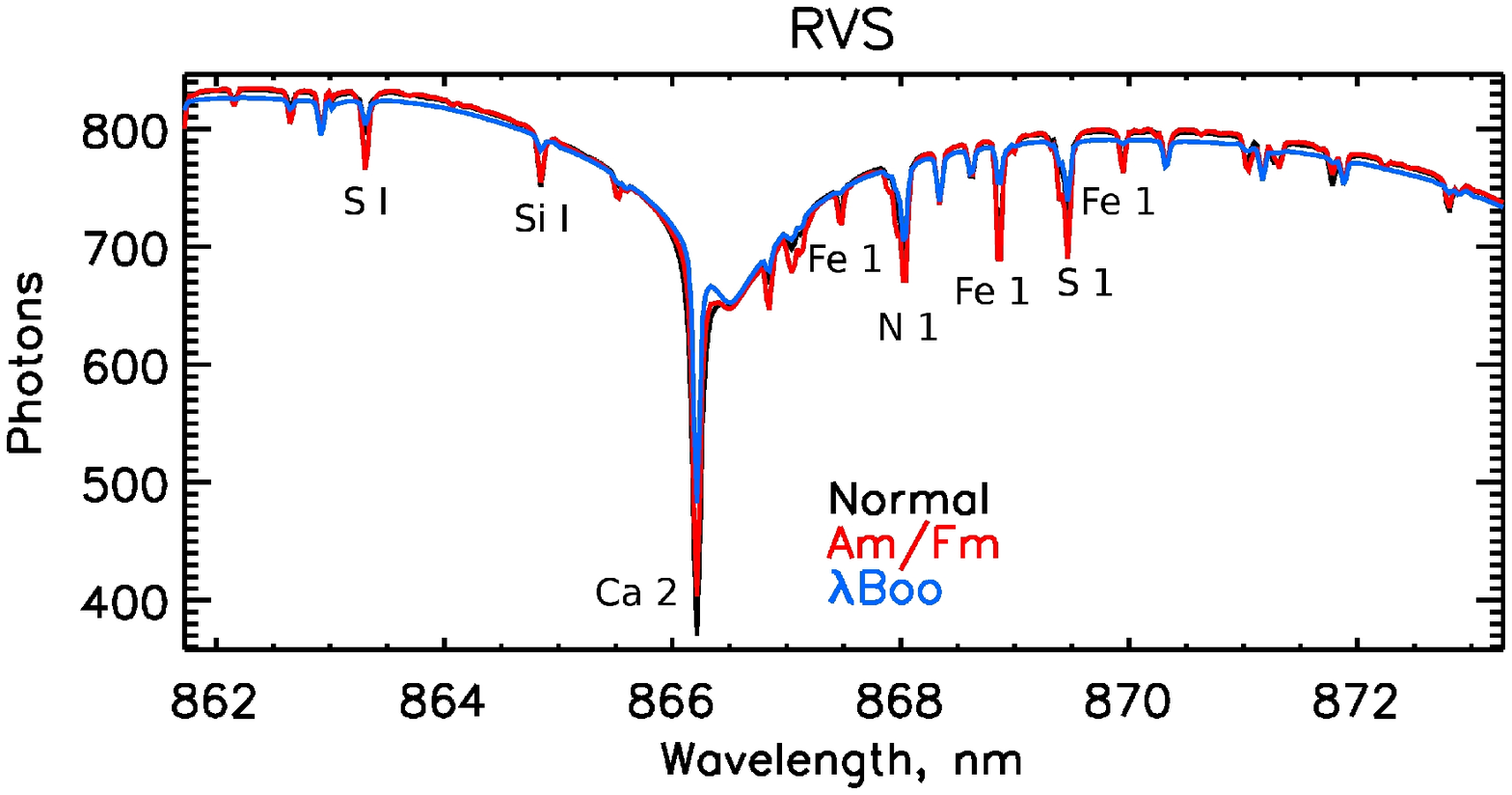}
 \caption{Same as in Fig.~\ref{fig:ap} but for Am/Fm and $\lambda$~Boo star models.}
 \label{fig:am}
\end{center}
\end{figure}

Having a suitable code for computing CP star spectra, it is now possible to
verify how does Gaia see CP stars. To simulate Gaia data we used Gaia Object Generator (GOG): a tool originally designed
to obtain catalogue data and main database data (including mission final data) for the Gaia satellite.
User source input specification were used allowing us directly feed GOG with high resolution
fluxes obtained by \textsc{LLmodels}. Calibration and spectral noises were ignored, mostly because their final models are
not yet strictly defined. The model grid of CP stars were computed taking into account characteristic chemistry 
and effective temperatures of following types of CP stars: Am/Fm, $\lambda$Boo, Ap, HgMn, He-weak and He-rich.

As an example, Fig.~\ref{fig:ap} illustrates the result of simulation of BP, RP, and RVS spectra of normal 
$T_{\rm eff}=8000$~K, $\log(g)=4.0$ star with solar composition and two Ap stars: non-magnetic and magnetic 
with assumed $10$~kG surface magnetic field. It is clearly seen the energy redistribution due to peculiar abundances
and magnetic field, and abnormally strong lines of REE in RVS spectrum in case of Ap stars. 
Thus, there is a chance to use both low resolution spectrophotometry and high resolution
spectroscopy to verify the type of peculiarity. The example of Am/Fm and $\lambda$~Boo models is presented in Fig.~\ref{fig:am}.

It follows from our analysis that using BP/RP spectra its possible to distinguish only between Ap and normal stars. 
Energy distribution of Am/Fm, $\lambda$Boo, HgMn, He-weak and He-rich stars as seen by Gaia are hardly different from that
of normal stars with the same  $T_{\rm eff}$ and $\log(g)$. However, in RVS spectra
different types of CP stars are well visible due to the presence of certain spectral features 
(like deep FeI and CaII lines in case of Am/Fm stars, REE elements in case of Ap stars, MnI/II features in case of HgMn 
stars, etc.).

Thus, using the BP/RP spectra can most likely help to see CP stars with only strong peculiarities causing substantial energy
redistribution from UV to IR and thus easily detectable. On the other hand, analysis of line strength indices or relative
parameters (like equivalent widths) in RVS spectra can be applied for all type of CP stars. In addition, after a standard 
procedure of Gaia's data processing (which will derive fundamental parameters of all observed stars), it will be possible
(based on RVS spectra) to derive extended astrophysical parameters like magnetic field strength, signatures of 
element stratification, starspots, etc.

In the framework of this analysis we carry out calculations of extensive grids of normal and selected CP groups model
atmospheres and fluxes, as well as development, testing and implementation of modified stellar parametrization algorithms
as applied to CP stars research.

During its $5$ years life time, Gaia will observe every object from $40$ to $250$ times depending upon their
individual positions on the sky. The Gaia time sampling is quite irregular, with gaps of typically a month. 
Still, the probability to recover the periods of strictly periodic signals is high, and this opens a possibility 
to study a variability of Ap and related stars.
The variability amplitude of magnetic CP stars depends on wavelength and is up to $\approx0.1$~mag.
For stars with $T_{\rm eff}>10\,000$~K the amplitude is lower for longer wavelength (decreasing amplitude from U to V), 
and with  $T_{\rm eff}<10\,000$~K variability in different filters may be in correlation and thus applying too wide filters
may result in very weak or no observed variations at all. The characteristic periods are from $0.5$~day to decades, and are
generally stable.

The preliminary estimate of detected variable Ap stars with Gaia is about $40\,000$ (lower limit) and
depends upon the Galaxy model used. This is achieved by the high photometric precision of Gaia in $G$-band ($350-1100$~nm)
which is $20$~mmag at $G=20$ and $\approx1$~mmag at $10<G<14$. In addition, the application of two separate BP and RP
photometers is a good point as it allows to study variability due to energy redistribution from UV to visual and IR.

On the other hand, the detection of roAp variability is a challenging task for Gaia since the characteristic amplitudes
are below $0.01$~mag and periods are between $6$ and $20$ minutes, however, there is a possibility to use data 
from every single CCD of Gaia detector separetely, thus allowing for finer time resolution which
is critical for roAp stars. 
For instance, \citet{mary} studied a model
of roAp star HR~3831 for which they were able to recover three periods assuming stable multiperiodic sinusoidal signal 
with $16$ frequencies (without noise). Later, \citet{varadi} investigated ZZ~Ceti stars resulting in 65\% of recovery 
of a period (multi-periodic nonlinear stable signal with $7$ frequencies, with noise at $G=18$~mag). Thus, the goal of using
Gaia data in the light of variability research is to detect and classify correctly roAp with 
some period(s) and amplitude(s) characteristics.

\begin{acknowledgements}
DS would like to acknowledge the support received from the Deutsche Forschungsgemeinschaft (DFG) Research Grant RE1664/7-1 and
IAU GA travel grant.

Personal thanks form DS to the GOG WEB administration team.
\end{acknowledgements}

{}

\end{document}